\documentclass{optica-article}

\journal{opticajournal} 

\articletype{Research Article}

\usepackage{lineno}

\usepackage{upgreek}
\usepackage{physics}

\usepackage{bm}
\usepackage{booktabs}

\usepackage{algorithm}
\usepackage{algpseudocode}
\usepackage{xcolor}
\usepackage[normalem]{ulem}

\begin{document}
\newcommand{\daan}[1]{{\color{purple} #1}}
\newcommand{\review}[1]{{#1}}
\newcommand{\remove}[1]{{\color{red} \sout{#1}}}

\title{Fast reconstruction of tensor tomographic X-ray scattering data for real-time applications}

\author{André Mesquita Antunes,\authormark{1,*}, Daniël Maria Pelt,\authormark{1} and Kees Joost Batenburg\authormark{1}}

\address{\authormark{1}Leiden Institute of Advanced Computer Science, Universiteit Leiden, Einsteinweg 55, 2333 CC Leiden, The Netherlands}

\email{\authormark{*}a.r.mesquita.fery.antunes@liacs.leidenuniv.nl} 


\begin{abstract*}

X-ray scattering tensor tomography reveals nanoscale structural orientation in 3D, but its reliance on slow iterative reconstruction limits real-time use. We introduce {an extension} of a direct reconstruction approach {for parallel-beam geometries} that computes algebraic filters approximating multiple iterative updates with a single filtering and back-projection step. By explicitly separating the tomographic projector from a view-dependent mixing operator, the method generalizes across tensor representations and modes of (small-angle) X-ray scattering measurement. Simulated and experimental results show that the approach approximates iterative reconstructions while reducing computation time by over an order of magnitude, realizing a reconstruction of {a 53x53x53x28 tensor volume in 1 second} on commercially-available hardware. The resulting speed and stability enable high-throughput analysis and open the door to real-time scattering-based imaging.

\end{abstract*}


\section{Introduction}

X-ray scattering tensor tomography is a class of imaging techniques that can measure high resolution (i.e., sub-voxel) structural information of scanned samples. These techniques are based on measuring the scattering of X-rays induced by sub-voxel sized sample structures, while rotating and tilting the sample between separate acquisitions to obtain three-dimensional information. Various methods for measuring scattering signals exist, including direct scanning-based methods \cite{Liebi2015,Schaff2015,Liebi2018} and full-field approaches with additional modulating optics \cite{Malecki2014,Kim2020,Lautizi2024b}, which we refer to in this article as different imaging modalities. Because of their ability to resolve sub-voxel structures, tensor tomography is used to study samples in various application fields, including biological tissues \cite{Georgiadis2021}, fibre-reinforced composites \cite{Kim2022b,Auenhammer2024}, and materials with a structural hierarchy \cite{Murer2021,Grunewald2024}, where orientation-dependent scattering provides insight into mechanical properties.

Recent improvements in detector technology, sample handling and source brightness have significantly reduced acquisition times \cite{Appel2025}. Together with improvements made towards fast and dynamic scanning protocols \cite{Buurlage2019}, these advances steer scattering-based tomography towards real-time imaging. In particular, real-time access to the scattering tensor would allow for imaging parameters (e.g. field of view) or other experimental conditions (e.g. applied mechanical load) to be immediately adjusted in response to the observed state of the sample, thus enabling feedback-driven experimental strategies and different kinds of advanced real-time visualizations. In order to fully realise this potential, the reconstruction procedure should take place in real time as well.


Reconstruction methods for conventional computed tomography (CT) fall into two main categories: direct and iterative. Iterative methods such as the simultaneous iterative reconstruction technique (SIRT) \cite{Kak2001} can produce accurate reconstructions even from limited data, but often require computation times that are too long for real-time settings \cite{Pelt2016,Lagerwerf2020,Buhrer2021}. Direct methods such as filtered back-projection (FBP) \cite{Kak2001} can produce reconstructions with significantly lower computation times, making them widely used for real-time reconstruction and visualization in conventional CT \cite{Buurlage2018,Lagerwerf2021,Graas2023}. Currently, to the best of our knowledge, there is no representation-independent direct method for tensor tomography, and iterative methods are typically used for reconstruction instead \cite{Kim2020,Nielsen2023,Carlsen2024,Lautizi2024a}. Although recent software developments have improved the speed of iterative tensor reconstruction \cite{mumott}, real-time applications of tensor tomography could benefit from the development of direct reconstruction methods.

In this paper, we {extend} a direct reconstruction method for X-ray scattering-based tensor tomography {to other acquisition modalities}, significantly reducing computation times comparatively to existing iterative methods. Our approach is based on computing algebraic filters that, when coupled with a single back-projection operation, approximate the output of iterative methods. The computed filters do not depend on the specific sample being scanned and can therefore be pre-computed for a specific set of acquisition parameters, enabling effective implementation in real-world settings. We {verify} the applicability of our method to scanning small-angle X-ray scattering tensor tomography (SASTT) \cite{Liebi2015,Liebi2018,Nielsen2023}, grating-based full-field tensor tomography (GITT) \cite{Kim2020,Kim2022a,Kim2022b}, and speckle-based dark-field tomographic imaging (SBTT) \cite{Lautizi2024a,Lautizi2024b}, demonstrating that a wide variety of state-of-the-art image acquisition methods can be combined with our technique to achieve real-time tensor reconstruction.

Our approach is based on similar methods for conventional CT \cite{Batenburg2012, Pelt2015}, which have proven effective for fast reconstruction of large real-world tomographic datasets \cite{Pelt2016} and for real-time observation of the internal dynamic processes of a scanned sample \cite{Buurlage2019}. {The demonstration of an algebraic filter reconstruction for tensor tomography was pioneered in Kim \textit{et al.} \cite{Kim2022a} in the context of GITT, but its extension to other tensor tomography modalities was left as an open avenue for future work. With linear forward models consolidated for various scanning and full-field modalities \cite{Kim2020,Nielsen2023,Lautizi2024a}, we demonstrate how to extend the algebraic filters reconstruction framework to parallel-beam SASTT and SBTT by introducing a mathematical abstraction that is compatible with those forward models. Since this abstraction does not require \textit{a priori} specification of either the physical scattering model or the representation of the scattering tensor, it enables the successful extension of the fast algebraic filters reconstruction method, as demonstrated by the experiments in this paper.} 

The paper is structured as follows. Section \ref{sec:PAPER1_background} establishes the mathematical framework of scattering tensor tomography and how it links to different experimental modalities. We approach the topic of reconstruction, introducing iterative methods and the concept of algebraic filters. Section \ref{sec:PAPER1_method} describes the proposed fast tensor tomography reconstruction algorithm and explains how it relates to different object representations. In Section \ref{sec:PAPER1_Results}, we investigate its performance on both synthetic and real data and discuss the results. Finally, Section \ref{sec:PAPER1_Conclusion} concludes with a summary of the work and a critical assessment of its implications.

\section{Background} \label{sec:PAPER1_background}
\subsection{Object representations}

In conventional X-ray CT, a three-dimensional description of a sample's local X-ray absorption is reconstructed from a set of radiographs, known as the tomographic projections. The reconstructed object, $\mathbf{x}$, is therefore typically represented as a discretized scalar volume, where to each voxel corresponds a single gray value. Conversely, in X-ray scattering-based tensor tomography, the local scattering strengths and 3D anisotropies are reconstructed from measurements across a set of tomographic angles. This scattering information reveals the local morphology and alignment of the nanostructures that cause the scattering itself. Within this framework, the reconstructed object is represented as a vector or tensor volume instead, where each voxel contains more than one scalar value to account for the comparatively higher complexity of the local scattering information.  

\begin{figure}[htbp]
    \centering
    \includegraphics[width=\linewidth]{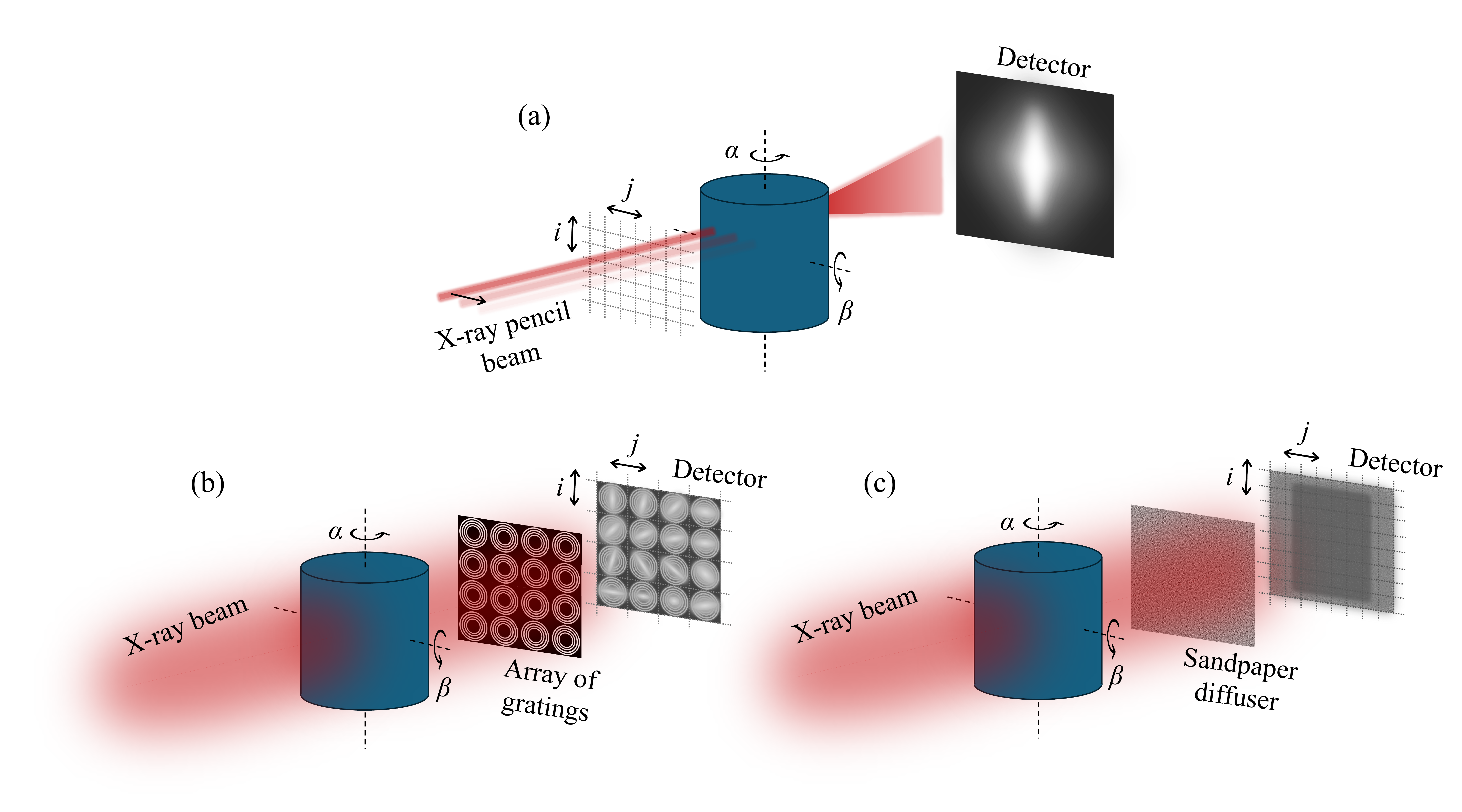}
    \caption{Simplified geometrical diagrams of three experimental variations of X-ray scattering-based tensor tomography. (a) Scanning small-angle X-ray scattering tensor tomography (SASTT): a very narrow X-ray beam raster-scans the sample, acquiring a $q$-resolved scattering pattern for each beam position $(i,j)$ and each set of tomographic ($\alpha$) and tilt ($\beta$) angles. The width of the beam determines the voxel size. (b) Full-field directional dark-field/grating interferometry tensor tomography (GITT) with a circular grating array: a wide X-ray beam illuminates the object and subsequent wavefront modulator comprised of an array of small diffraction gratings. The grating periodicity determines the reconstruction voxel size. (c) Full-field speckle-based tensor tomography (SBTT): similar to (b), but with the grating modulator swapped with a set of sandpaper diffusers which are repeatedly translated for each set of tomographic and tilt angles. The voxel size is determined by the choice of analysis window of the detected signals.}
    \label{fig:PAPER1_diagrams_modalities}
\end{figure}

Different experimental setups for acquiring small-angle X-ray scattering signals exist, and each of these can motivate modelling the physics of X-ray scattering differently. The choice of forward model implicitly determines the allowed representations of its input, the tensor tomogram. However, when the physical model allows it, a different representation of the object may be used to highlight certain features or symmetries of the scattering tensor. The three imaging modalities that {are} considered in this article are described in Fig. \ref{fig:PAPER1_diagrams_modalities}. These modes of acquisition all assume a tomographic parallel-beam geometry and measure scattering anisotropy per detector element. 

\begin{figure}[htbp]
    \centering
    \includegraphics[width=\linewidth]{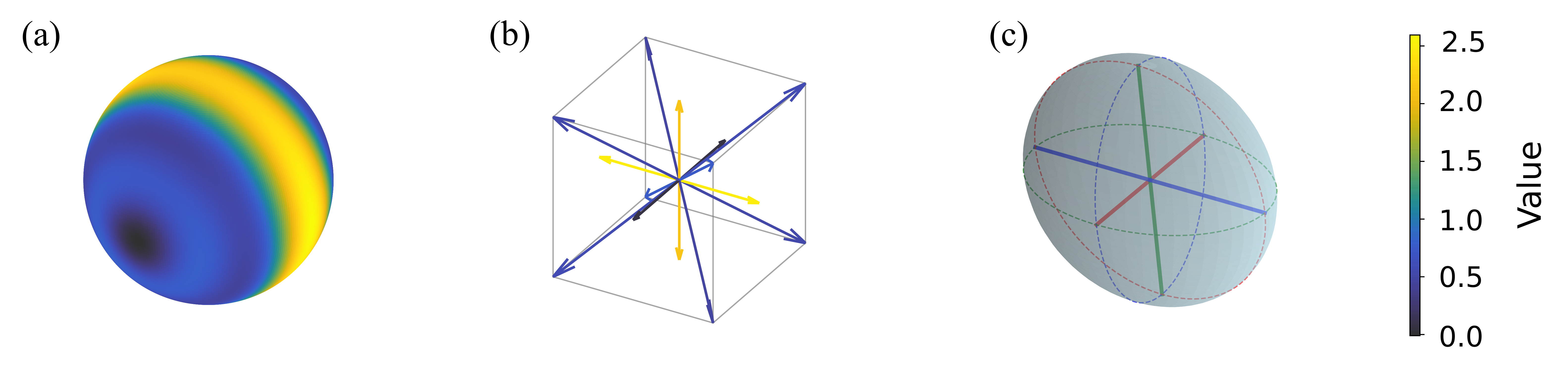}
    \caption{Example visualization of three possible representations of the same scattering tensor of a single voxel. (a) Smooth spherical function depiction, resulting from a finite set of coefficients of a spherical harmonic expansion. (b) Three-dimensional directional dark-field values, visualised as the colour of a finite set of oriented arrows. (c) Second angular moment tensor, visualized as the ellipsoid described by its corresponding quadratic form.}
    \label{fig:PAPER1_rsm_representation_diagram.pdf}
\end{figure}

Some models that have been used for the experimental modalities described in Fig. \ref{fig:PAPER1_diagrams_modalities} reconstruct a quantity on the unit sphere per voxel. Recent approaches in the context of SASTT (Fig. \ref{fig:PAPER1_diagrams_modalities}(a)) opt for a finite set of real-valued spherical harmonics to approximate the reciprocal-space map within each voxel, and reconstruct the expansion coefficients \cite{Nielsen2023}. The SASTT model description is also compatible with choosing a set of Gaussian kernels to approximate the spherical function, or with directly sampling and dealing only with certain points on the sphere (or directions, if this function is centrosymmetric). This is also the approach of the first works surrounding GITT as it is described in Fig. \ref{fig:PAPER1_diagrams_modalities}(b) \cite{Kim2020,Kim2022a,Kim2022b}, although compatibility with a continuous representation of the scattering tensor has also been demonstrated \cite{Wieczorek2016}. More recent work centered on the modelling of the full-field modalities of Fig. \ref{fig:PAPER1_diagrams_modalities}(b) and Fig. \ref{fig:PAPER1_diagrams_modalities}(c) represents the local scattering content per voxel as a symmetric, positive-semidefinite rank-2 tensor \cite{Lautizi2024a,Lautizi2024b}. Fig. \ref{fig:PAPER1_rsm_representation_diagram.pdf} shows an example visualization of three different representations of a scattering tensor.

\subsection{Modelling scattering tensor tomography}

Despite the different possible representations of the tensor volume $\mathbf{x}$, the X-ray scattering tensor tomography problems underlying the imaging modalities of Fig. \ref{fig:PAPER1_diagrams_modalities} can all be modelled as a linear system:

\begin{equation}
    \mathbf{A}\mathbf{x} = \mathbf{\overline{W}Yx} = \mathbf{b} \, ,
    \label{eq:PAPER1_WYx=b}
\end{equation}

\noindent where the object $\mathbf{x} \in \mathbb{R}^{N\cdot C}$ is a tensor volume of $N$ voxels and $C$ tensor components, $\mathbf{b} \in \mathbb{R}^{M\cdot S}$ is the measured scattering anisotropy data for $M$ projections and $S$ anisotropy channels, and the linear operator $\mathbf{A} \in \mathbb{R}^{(M\cdot S)\times(N\cdot C)}$ describes the complete forward model.

For the linearized models commonly used in the context of the modalities of Fig. \ref{fig:PAPER1_diagrams_modalities}, $\mathbf{A}$ can be split into a tensor-compatible tomographic projector $\mathbf{\overline{W}}$ that applies the conventional projector $\mathbf{W}$ to each tensor component independently, and a linear operator $\mathbf{Y}$ that contains the physical information about how X-rays scatter as they traverse the sample. In practical terms, $\mathbf{Y}$ links the representation of the tensor to be reconstructed to the scattering information collected in the measurements $\mathbf{b}$. Since the local scattering properties of the sample depend on the relative orientation between the incoming beam and the scattering structures, $\mathbf{Y}$ must be view-dependent. In this paper, we refer to $\mathbf{Y}$ as the mixing operator, since it determines which parts of the scattering tensor contribute to which anisotropy measurement. 


Different experimental modalities of X-ray scattering-based tensor tomography motivate modelling the problem, and consequently $\mathbf{Y}$, differently. In the context of SASTT (Fig. \ref{fig:PAPER1_diagrams_modalities}(a)), the full forward model sums slices of 3D reciprocal space maps (which are the multi-dimensional quantities to be reconstructed) along the path of the narrow scanning beam \cite{Nielsen2023}. In the full-field approaches with wavefront-modulating elements (Fig. \ref{fig:PAPER1_diagrams_modalities}(b) and \ref{fig:PAPER1_diagrams_modalities}(c)), the commonly used models sum projections of the tensor quantities onto the plane perpendicular to the beam direction \cite{Kim2020,Lautizi2024a}.

The choice of object representation may influence the quality or determine the physical meaning of the reconstruction. However, within the linearized model descriptions of the modalities in Fig \ref{fig:PAPER1_diagrams_modalities}, it does not influence the applicability of our proposed algorithm.

\subsection{Tomographic reconstruction}

Tomographic reconstruction algorithms generally fall into one of two categories: direct or iterative. Direct algorithms, in the case of conventional CT, come from the discretization of an analytic inversion formula that is derived from the forward model \cite{Kak2001}. As such, they are rather fast, but also sensitive to noise and prone to discretization artifacts, especially in limited-data scenarios \cite{Pelt2016,Bicer2017}. The direct reconstruction algorithm for a parallel-beam geometry is called filtered back-projection (FBP) \cite{Hansen2021}, since it consists of a filtering operation done over the projection data, followed by the application of the adjoint operator $\mathbf{W}^\top$ (the back-projection). This algorithm can be summarized as 

\begin{equation}
    \mathbf{x}_\mathrm{recon} = \mathrm{FBP}(\mathbf{b}, \mathbf{h}) = \mathbf{W}^\top \mathbf{C}_{\mathbf{h}} \mathbf{b}\, ,
    \label{PAPER1_FBP_CT}
\end{equation}

\noindent where $\mathbf{C}_{\mathbf{h}}$ describes the convolution of a ramp filter $\mathbf{h}$ with the projection data $\mathbf{b}$.


Iterative reconstruction algorithms, on the other hand, only require knowledge of the forward and adjoint operators, without the need of problem-specific information. These algorithms are also typically better than direct methods at handling noisy and limited-view data, often yielding higher quality reconstructions \cite{Pelt2016, Zhu2019}. However, since they function by successive application of the forward and adjoint operators to some initial guess of $\mathbf{x}$ until convergence is reached, they are also slower than direct methods. In conventional CT, where $\mathbf{W}$ is the complete forward operator, an iterative reconstruction algorithm such as the Landweber method \cite{Hansen2021} may be expressed as an update equation:

\begin{equation}
    \mathbf{x}^{(k)} = \mathbf{x}^{(k-1)} + \alpha \mathbf{W}^\top(\mathbf{b} - \mathbf{W}\mathbf{x}^{(k-1)})\, ,
    \label{eq:PAPER1_landweber_iteration}
\end{equation}

\noindent where $\mathbf{x}^{(k)}$ is the reconstruction solution after $k$ iterations of the algorithm, and $\alpha$ is a parameter that conditions its stability and speed of convergence.

\subsection{Algebraic filters}

The concept of algebraic filters originated in the field of conventional CT as a way of combining the high quality reconstructions of iterative methods and the speed of FBP \cite{Batenburg2012}. The fundamental underlying idea is to approximate the operations of an iterative algorithm by an FBP-like set of operations (filtering and back-projection) with the goal of producing a higher quality reconstruction in less time. 

The algebraic filter method relies on the assumption that the reconstruction should be shift-equivariant with respect to the projection data, i.e., translating the object should result in a similar displacement of its X-ray projections. Consequently, any closed formulation of the reconstruction value of a single voxel in the volume (e.g., the central voxel) immediately extends to an inversion formula for the full volume based on translating the projections, which can be efficiently implemented through a combination of convolutions and a back-projection.

 Given the simple recurrent formulation of an iterative algorithm such as the Landweber method of eq. \eqref{eq:PAPER1_landweber_iteration}, the conventional CT solution at the $k$th iteration may also be expressed as an application of a reconstruction operator $\mathbf{Q}$ \cite{Pelt2015}:

\begin{equation}
    \mathbf{x}^{(k)} = \alpha \left[\sum_{i=0}^{k-1} (\mathbf{I} - \alpha\mathbf{W}^\top\mathbf{W})^i \right]\mathbf{W}^\top \mathbf{b} = \mathbf{Q}\mathbf{b} \, ,
    \label{eq:PAPER1_landweber_Q}
\end{equation}

\noindent where the successive applications of the forward and back-projections are all enclosed in $\mathbf{Q}$. Then, if $\mathbf{Q}$ is approximately shift-equivariant, it may be expressed as a filtering (convolution) operation followed by a back-projection acting on the projections:

\begin{equation}
    \mathbf{x}_\mathrm{recon} = \mathbf{Q} \mathbf{b} \sim \mathrm{FBP}(\mathbf{b}, \mathbf{q}) = \mathbf{W}^\top \mathbf{C}_{\mathbf{q}}\mathbf{b}\, ,
    \label{eq:PAPER1_reconstruction_matrix}
\end{equation}

\noindent where $\mathbf{q}$ is called the algebraic filter. {Note that algebraic filter methods depend on the ability to express the reconstruction operation as a single linear shift-equivariant operator $\mathbf{Q}$, which is possible for the Landweber method but might not be possible for other iterative reconstruction methods (e.g., SIRT and CGLS).}

{The algebraic} filter can be determined by computing the impulse response of $\mathbf{Q}$ in volume space:

\begin{equation}
    \mathbf{q} = \mathbf{Q}^\top \bm{\updelta}^\mathrm{central} \, ,
    \label{eq:PAPER1_impulse_CT}
\end{equation}

\noindent where $\bm{\updelta}^{\mathrm{central}}$ is a scalar volume with only one non-zero entry, the central voxel, which is set to 1.

Once an algebraic filter is computed, the FBP-like reconstruction of equation \eqref{eq:PAPER1_reconstruction_matrix} replaces the multiple applications of $\mathbf{W}$ and $\mathbf{W}^\top$ with a single convolution and back-projection, making this method orders of magnitude faster than the original iterative reconstruction. Furthermore, the same filter can be re-used for any other sample as long as the acquisition geometry remains the same (i.e., the choice of volume discretization and angle sampling), enabling high-throughput scanning and real-time visualization.


\section{Method} \label{sec:PAPER1_method}
In this section, we describe the main contribution of this paper: {a representation-independent} framework for computing algebraic filters for {parallel-beam} scattering-based tensor tomography, and achieving a fast FBP-like reconstruction with their usage.
The proposed approach is similar to the one in Kim \textit{et al.} \cite{Kim2022a}, where computing algebraic filters for tensor tomography is achieved for a specific mixing operator and object representation associated with grating-based dark-field tensor tomography {(GITT)}. Here, we extend its applicability to other representations and definitions of $\mathbf{Y}$, and {extend} its {mathematical} framework {to enable the automatic determination of} a model-appropriate step size $\alpha$ using a power method.

\subsection{Algebraic filters for tensor tomography} \label{subsec:PAPER1_methods_aftt} 

Given the presence of the view-dependent operator $\mathbf{Y}$ in the forward model of scattering-based tensor tomography (eq. \eqref{eq:PAPER1_WYx=b}), straightforward application of FBP is no longer possible. In fact, FBP can only be used in a small subset of scattering-based tomography problems where scattering can be assumed to be isotropic \cite{Schaff2015}, i.e., whenever the measured scattering response can be assumed to be unique for each and every voxel independently of the tomographic view. Currently, and to the best of our knowledge, the absence of a representation-independent inversion formula for scattering-based tensor tomography means that iterative reconstruction algorithms are commonly used instead \cite{Malecki2014,Gao2019,Nielsen2023,Kim2020,Lautizi2024a}, despite their computational cost. 

Algorithms like the Landweber method are operator-agnostic, which means they can be adapted for tensor reconstruction simply by replacing the tomographic projector $\mathbf{W}$ with the complete forward operator of the tensor tomography model $\mathbf{A} = \mathbf{\overline{W}}\mathbf{Y}$ of equation \eqref{eq:PAPER1_WYx=b}:

\begin{equation}
    \mathbf{x}^{(k)} = \mathbf{x}^{(k-1)} + \alpha \mathbf{A}^\top(\mathbf{b} - \mathbf{A}\mathbf{x}^{(k-1)})\, ,
    \label{eq:PAPER1_landweber_iterations_tensor}
\end{equation}

\noindent where $\mathbf{x}^{(k)}$ is now the tensor volume reconstruction after $k$ Landweber iterations. {The Landweber method will converge to the least-squares fit of the data, similarly to other iterative algorithms typically used for tensor tomography reconstruction \cite{Lautizi2024a, Nielsen2023}.} Just as in eq. \eqref{eq:PAPER1_landweber_Q}, a tensor reconstruction operator $\mathbf{Q}$ can be formulated: 

\begin{equation}
    \mathbf{x}^{(k)} = \alpha \left[\sum_{i=0}^{k-1} (\mathbf{I} - \alpha\mathbf{A}^\top\mathbf{A})^i \right]\mathbf{A}^\top \mathbf{b} = \mathbf{Q}\mathbf{b} \, .
    \label{eq:PAPER1_landweber_Q_tensor}
\end{equation} 

{For the algebraic filters framework to be applicable, the reconstruction operator $\mathbf{Q}$ must be linear and approximately shift-equivariant. To ensure shift-equivariance of the operations contained in $\mathbf{Q}$, both $\mathbf{\overline{W}}$ and $\mathbf{Y}$ need to be shift-equivariant. By noticing that the tomographic view angle is the same for all voxels in a parallel-beam geometry, and that $\mathbf{Y}$ is independent of the voxel index of $\mathbf{x}$, we can conclude that the mixing operator $\mathbf{Y}$ is shift-equivariant by construction. Therefore, the shift-equivariance  of $\mathbf{Q}$ depends entirely on the approximate shift-equivariance of the tomographic projector $\mathbf{\overline{W}}$, which is well-established for parallel-beam geometries \cite{Pelt2015}.}

In order to simplify the problem of computing algebraic filters for tensor tomography, it helps to view the forward model as a collection of intertwined scalar tomography problems

\begin{equation}
    \sum_{c=1}^C \mathbf{W}\mathbf{Y}_c \mathbf{x}_c = \mathbf{b}\, , 
    \label{PAPER1_WYkxk=b}
\end{equation}


\noindent where the index $c$ runs over the number of tensor components $C$ of $\mathbf{x}$. This framework explicitly connects the usual tomographic operator $\mathbf{W}$ to the scalar volume $\mathbf{x}_c$, which motivates the search for one algebraic filter per tensor component. The computation of a set of algebraic filters for tensor tomography then begins with building up the reconstruction operator $\mathbf{Q}$ for the Landweber method as in \eqref{eq:PAPER1_landweber_Q_tensor}. In this case,

\begin{equation}
    \mathbf{Q}^{(k)} = \alpha \left[\sum_{i=0}^{k-1}(\mathbf{I}-\alpha \mathbf{A}^\top \mathbf{A})^i\right]\mathbf{A}^\top
    \label{eq:PAPER1_Q_landweber_tt}
\end{equation}

\noindent is defined for a given number of $k$ iterations.

{An appropriate Landweber relaxation parameter $\alpha$ can be computed irrespectively of the exact definition of $\mathbf{A}$. This is a key advantage of the representation-independent formulation, automatically adapting $\alpha$ to the physical model and tensor representation embedded in $\mathbf{Y}$ and $\mathbf{x}$. The Landweber parameter $\alpha$ can be computed by estimating} the largest {eigenvalue $\lambda_\mathrm{max}$ of $\mathbf{A}^\top\mathbf{A}$} after a few power iterations on the back-projected data, if available, or on a randomly-initialized volume \cite{Bertero1998,Golub2012}{. A choice of $\alpha$ close to the theoretical upper bound for which convergence is guaranteed ($2/\lambda_{\mathrm{max}}$)} ensures faster convergence of the iterative method. A smaller number of iterations, in turn, reduces the accumulating errors of the shift-equivariance approximation caused by the successive applications of $\mathbf{A}$ and $\mathbf{A}^\top$ and will result in a better quality algebraic filter reconstruction.

Then, by noting that we can consider a separate reconstruction operator for each tensor component $c$:

\begin{equation}
    \mathbf{x}_c = (\mathbf{Q}\mathbf{b})_c  \, , 
    \label{eq:PAPER1_xc_from_Qb}
\end{equation}

\noindent we can then probe its component-specific impulse response in order to find the corresponding algebraic filter:

\begin{equation}
    \mathbf{q}_c = (\mathbf{Q}^\top \bm{\updelta}^{\mathrm{central}(c)})_c\, ,
    \label{eq:PAPER1_impulse_TT}
\end{equation}

\noindent where $\bm{\updelta}^{\mathrm{central}(c)}$ is a tensor volume with the same structure as $\mathbf{x}$. Similar to equation \eqref{eq:PAPER1_impulse_CT}, its only non-zero entry is the central voxel of $\bm{\updelta}_c$, set to 1. Further details about the pre-computation of a set of algebraic filters for scattering-based tensor tomography can be found in Algorithm \ref{alg:PAPER1_compute_algfilters}. 

\begin{algorithm}[htbp]
\caption{Algebraic filter computation for X-ray scattering tensor tomography}
\label{alg:PAPER1_compute_algfilters}
\begin{algorithmic}[1]
\Require Tensor-compatible tomographic projector $\mathbf{\overline{W}}$ and adjoint $\mathbf{\overline{W}}^{\top}$; mixing operator $\mathbf{Y}$; number of (Landweber) iterations $k$.
\Ensure Set of tensor tomography algebraic filters $\mathbf{q}=\{\mathbf{q}_c\}_{c=1}^{C}$.
\State Define: $\mathbf{A} \gets \mathbf{\overline{W}Y}$ and its adjoint $\mathbf{A}^{\top}$.

\Statex Estimate $\lambda_{\mathrm{max}}({\mathbf{A}^\top}\mathbf{A})$ with $n_{\mathrm{power}} = 8$ power iterations:
\State Initialize tensor volume $\mathbf{v} \gets \mathrm{random}$.
\For{$\_=1$ to $n_\mathrm{power}$}:
    \State Compute auxiliary $\mathbf{w} \gets \mathbf{A}^\top\mathbf{A}\mathbf{v}$.
    \State Update $\lambda_{\mathrm{max}} \gets \mathbf{v}\cdot\mathbf{w}$.
    \State Normalize and update for next step $\mathbf{v} \gets \mathbf{w}/||\mathbf{w}||$.
\EndFor

\State Set the relaxation parameter $\alpha \gets 1.9/\lambda_{\mathrm{max}}$. {\Comment{Choose a numerator close to but under 2}}
\State Define: $\mathbf{Q} = \mathbf{Q}^{(k)}$ and corresponding adjoint $\mathbf{Q}^\top$ according to eq. \eqref{eq:PAPER1_Q_landweber_tt}. 
\For{$c=1$ to $C$}
    \State Form $\bm{\updelta}^{\mathrm{central}(c)}$: a tensor volume of zeros with the central voxel of component $c$ set to $1$.
    \State Compute $\mathbf{q}_c \gets \bigl(\mathbf{Q}^{\top}\bm{\updelta}^{\mathrm{central}(c)}\bigr)_c$.
\EndFor
\State \Return $\mathbf{q} = \{\mathbf{q}_c\}_{c=1}^{C}$

\end{algorithmic}
\end{algorithm}

To obtain the $c$th component of the reconstruction, we convolve the corresponding algebraic filter with each image channel $s$ of the measurements $\mathbf{b}$ and back-project all channels jointly:

\begin{equation}
    \mathbf{x}_c \sim \mathrm{FBP}(\mathbf{b}, \mathbf{q}_c) = \mathbf{W}^\top \sum_{s=1}^S (\mathbf{C}_{\mathbf{q}_c} \mathbf{b})_s \, .
    \label{eq:PAPER1_AF_tensor}
\end{equation}

This process is repeated for each remaining tensor component until the full tensor object $\mathbf{x}$ has been retrieved. The algorithm of the FBP-like reconstruction using the pre-computed algebraic filters is described in Algorithm \ref{alg:PAPER1_reconstruct_algfilters}.

\begin{algorithm}[htbp]
\caption{Algebraic filter reconstruction for X-ray scattering tensor tomography}
\label{alg:PAPER1_reconstruct_algfilters}
\begin{algorithmic}[1]
\Require Scattering anisotropy data $\mathbf{b}=\{\mathbf{b}_s\}_{s=1}^{S}$; set of tensor tomography algebraic filters $\mathbf{q}=\{\mathbf{q}_c\}_{c=1}^{C}$; conventional tomographic back-projection operator $\mathbf{W}^\top$.
\Ensure Reconstructed tensor volume $\mathbf{x}=\{\mathbf{x}_c\}_{c=1}^{C}$.

\For{$c=1$ to $C$}
    \State Filter $\mathbf{b}$ with $\mathbf{q}_c$ and isolate anisotropy channel: $\tilde{\mathbf{b}}_{c,s} \gets (\mathbf{C}_{\mathbf{q}_c} \mathbf{b})_s \quad$ for $s=1,\ldots,S$.

    \State Back-project jointly: $\mathbf{x}_c \gets \mathbf{W}^{\top}\!\left(\sum_{s=1}^{S} \tilde{\mathbf{b}}_{c,s}\right)$ \Comment{Eq. \eqref{eq:PAPER1_AF_tensor}}
\EndFor
\State \Return $\mathbf{x}$
\end{algorithmic}
\end{algorithm}

\subsection{Implementation} \label{subsec:PAPER1_Implementation}

Our implementation of the complete forward model of eq. \eqref{eq:PAPER1_WYx=b} incorporates the GPU-accelerated, tensor-compatible tomographic projector $\mathbf{\overline{W}}$ from the \textit{mumott} library \cite{mumott} with array instantiations of the various mixing operators $\mathbf{Y}$ on the GPU using the \textit{cupy} package \cite{cupy}. Due to the intractable size of array implementations of the $\mathbf{Q}$ and $\mathbf{A}$ (and corresponding adjoints), we implement them as successive applications of matrix-vector operations. The mixing operator for the spherical harmonic representation of the scattering tensors {was} retrieved directly from \textit{mumott}'s basis set utilities \cite{mumott}, whereas the other two mixing operators (for the directional and tensor representations) were {implemented} with knowledge of the acquisition geometry. The mixing operator for the directional representation was constructed directly from eq. (3) of Kim \textit{et al.} \cite{Kim2022a}, and that of the tensor representation was computed according to the supplementary material of Lautizi \textit{et al.} \cite{Lautizi2024a}, by considering the scattering tensors as covariance matrices of three-dimensional multivariate Gaussian distributions and marginalizing them along the direction of the beam for each angle.

For the algebraic filters reconstruction, a parallel implementation on the CPU of the convolution operations was used, executing the exact same operations of \textit{scipy}'s FFT-based convolution algorithm \cite{scipy}. All experiments were executed on a workstation equipped with a NVIDIA GeForce RTX 4070 Super GPU and an Intel Core i5-14600KF (14-Core) CPU. More details about implementation, including a notebook with the complete simulation and reconstruction pipeline for the three tensor representations of the M-shaped phantom will be made available upon publication.

\section{Results} \label{sec:PAPER1_Results}

\subsection{Simulation study}

\begin{figure}
    \centering
    \includegraphics[width=\linewidth]{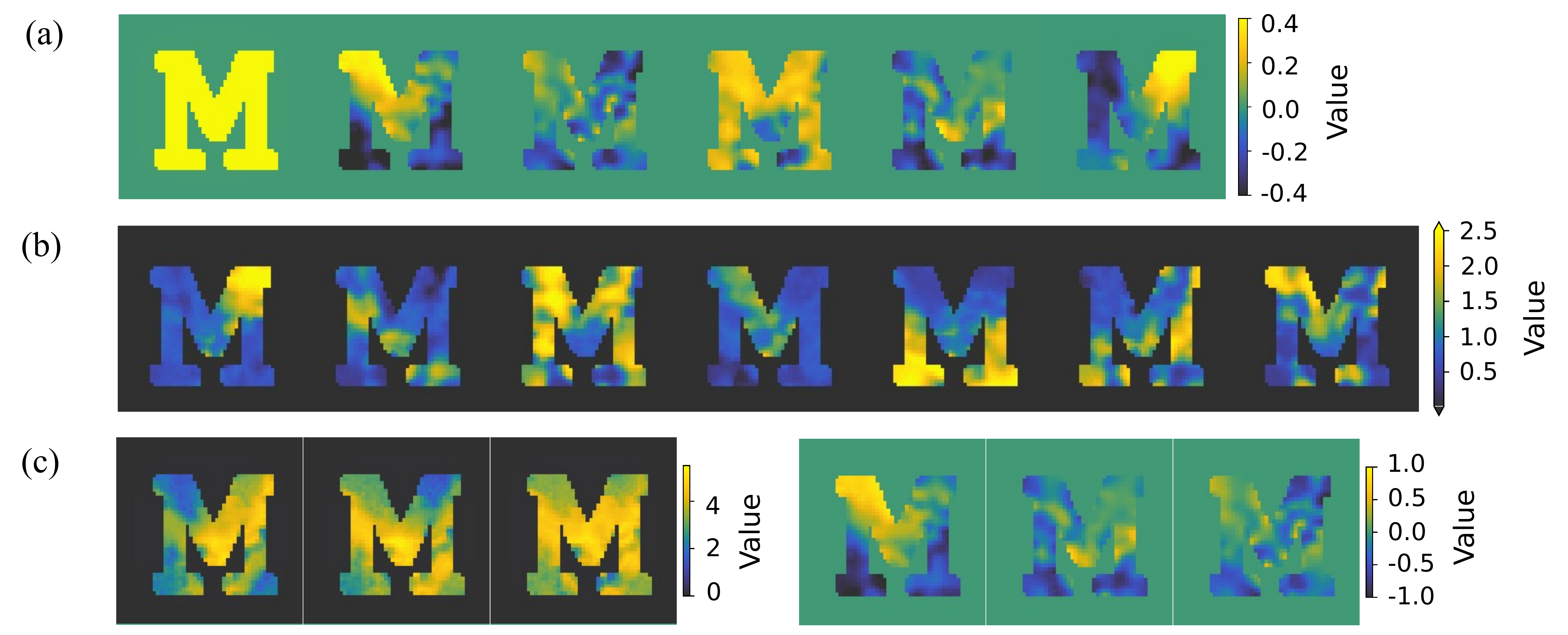}
    \caption{Visualization of example representations of the central slice of the synthetic M-shaped tensor phantom from Nielsen \textit{et al.} \cite{Nielsen2023,M_dataset}. (a) Values of the first six coefficients in a spherical harmonic expansion with even-order coefficients up to $\ell=6$. (b) Directional values retrieved from the evaluated expansion in (a) across the seven sampled directions portrayed in Fig. \ref{fig:PAPER1_rsm_representation_diagram.pdf}(b). (c) Six unique components of the scattering tensor defined in eq. \eqref{eq:PAPER1_second_order_moment}, using different scales for the diagonal (left) and off-diagonal (right) entries.}
    \label{fig:PAPER1_M_phantom_representations}
\end{figure}

In order to demonstrate and test our method, we perform a simulation study with the M-shaped tensor phantom from Nielsen \textit{et al.} \cite{Nielsen2023, M_dataset}. This synthetic sample consists of a 50x50x50-voxel volume, where each voxel stores the values of 28 coefficients in a spherical harmonic expansion representation of the scattering tensor. Fig. \ref{fig:PAPER1_M_phantom_representations}(a) shows the values of each coefficient for a central slice through the phantom volume. 
By evaluating and sampling the resulting spherical function over a set of directions, we can modify the scattering tensor representation to one compatible with the usual model descriptions of GITT (see Fig. \ref{fig:PAPER1_M_phantom_representations}(b)).
A symmetric, positive-semidefinite rank-2 tensor representation of the local scattering information is also retrieved from the (unnormalized) second-order spherical moment of the resulting spherical function $f(\hat{\mathbf{n}})$ per voxel:

\begin{equation}
    T = \int_0^{4\pi} f(\hat{\mathbf{n}}) \hat{\mathbf{n}} \otimes \hat{\mathbf{n}} \, \dd \Omega \, ,
    \label{eq:PAPER1_second_order_moment}
\end{equation}

\noindent where $\hat{\mathbf{n}} \otimes \hat{\mathbf{n}}$ is the outer product of the coordinate unit vector with itself. With a global isotropic shift applied to the evaluated spherical functions, we guarantee that the resulting tensors are positive-semidefinite, as parametrized in the model of Lautizi \textit{et al.} \cite{Lautizi2024a}, without compromising the functions' gradients. A visualization of the tensor entry values for this test phantom is shown in Fig. \ref{fig:PAPER1_M_phantom_representations}(c). 
{As a general pre-processing step} to mitigate artefacts at the edges of the reconstruction volume, where the shift-equivariance assumption breaks down, the reconstruction volume is padded with zeros to make sure the actual sample reconstruction is not influenced by this effect. {In practice, this steps amounts to making sure the projected volume completely fits within the projection, leaving a few empty pixels (at least 2) between the edge of the projected volume and the image boundary.} The amount of zero-padding is also chosen such that the reconstruction grid has an unequivocal central voxel, i.e., that its sides are composed of an odd number of voxels.

\begin{figure}
    \centering
    \includegraphics[width=\linewidth]{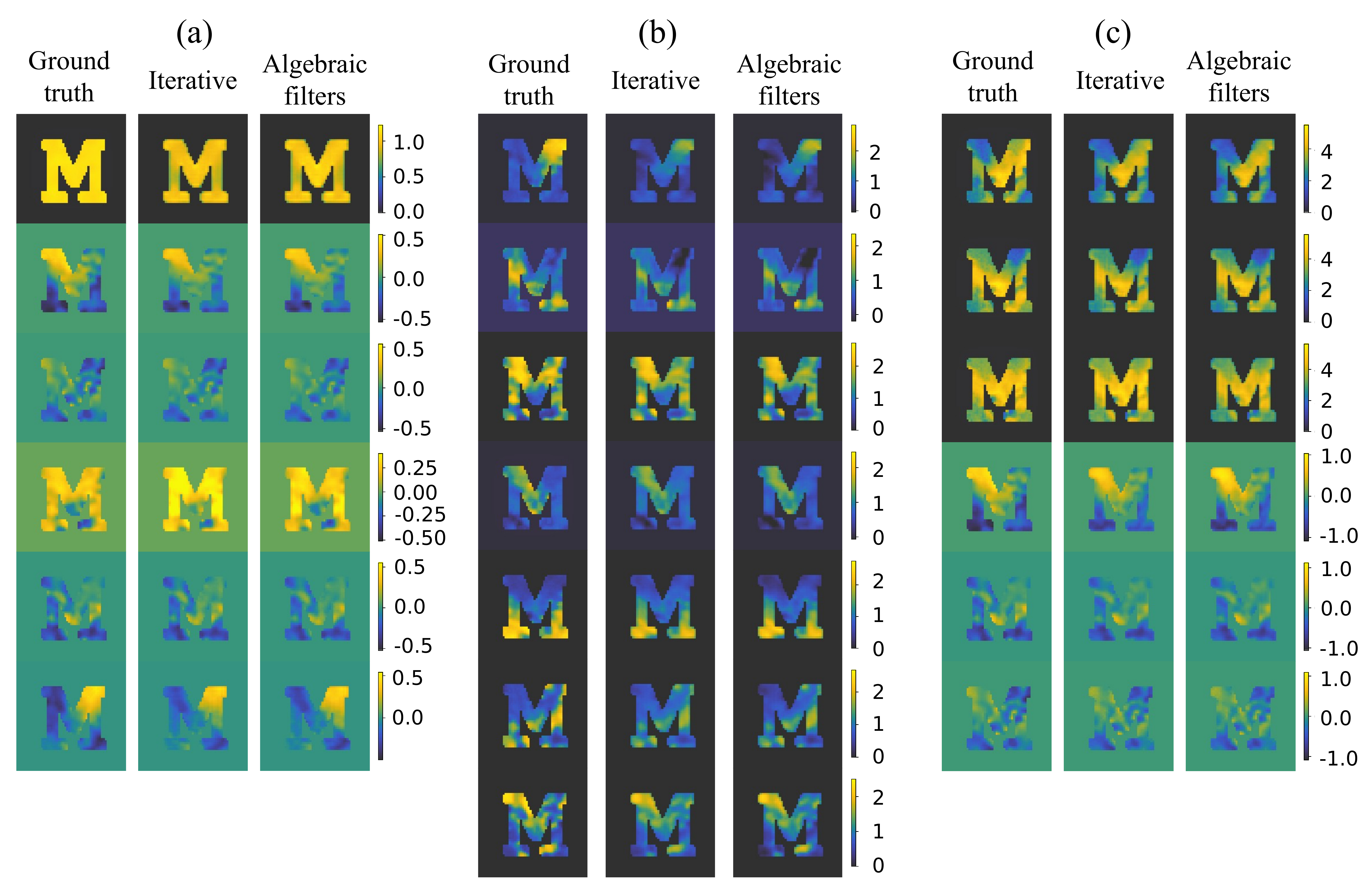}
    \caption{Visualization of the central slice of the M-shaped phantom compared to both iterative and algebraic filters reconstructions for three different scattering tensor representations and corresponding $\mathbf{Y}$ formulations, with $k = 50$ iterations. (a) First six coefficients of a spheric harmonic expansion representation; (b) seven sampled directions; (c) six unique tensor entries.}
    \label{fig:PAPER1_M_reconstructions}
\end{figure}

To validate the proposed approach, we start by simulating projection data of the M-shaped tensor phantom. For each of the three tensor representations and corresponding mixing operators considered in Fig. \ref{fig:PAPER1_rsm_representation_diagram.pdf}, we reconstruct the projections using both the Landweber iterative method and the proposed algebraic filters method with $k = 50$ iterations and $\alpha$ estimated in the same way across representations (see section \ref{subsec:PAPER1_methods_aftt}). Fig. \ref{fig:PAPER1_M_reconstructions} shows the central slice of a few select tensor component reconstructions against the ground truth for each object representation. 

The quality of the reconstructions is assessed on a masked region that restricts the error metric computations to the true extent of the 'M' volume. We compute the mean squared error (MSE) between the Landweber iterative reconstruction and the algebraic filters reconstructions, and the MSE of each reconstruction with respect to the ground truth. The former quantifies how well the proposed method approximates the iterative reconstruction it aims to reproduce.

\begin{figure}
    \centering
    \includegraphics[width=\linewidth]{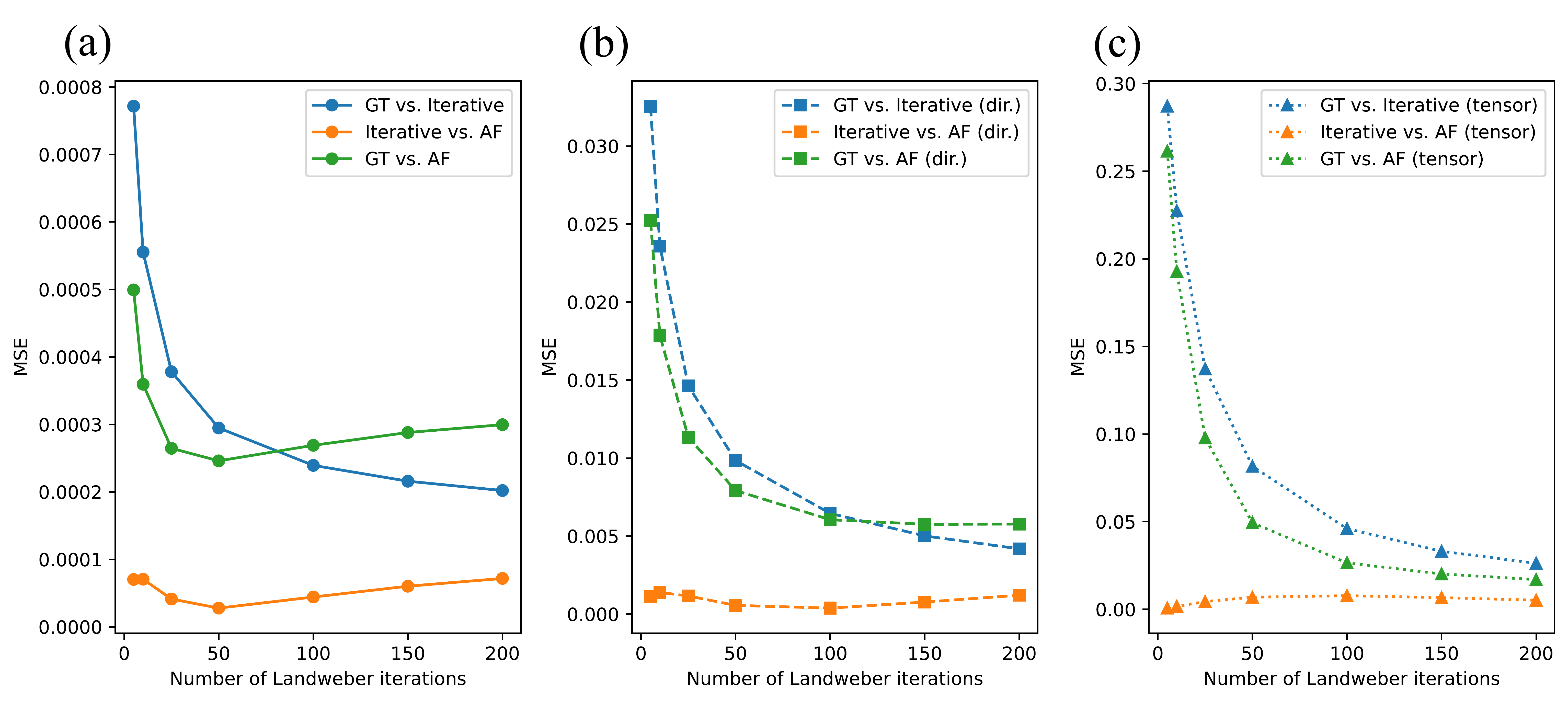}
    \caption{(a-c) Mean squared error (MSE) of the iterative Landweber reconstruction relative to the ground truth (GT) (blue curves), MSE of the algebraic filter reconstruction (AF) relative to the iterative result (orange curves), and MSE of the algebraic-filter reconstruction relative to the ground truth (green curves), shown as a function of the number of iterations and for each of the three considered object representations: (a) spherical harmonic representation (no abbreviation), (b) directional representation (dir.), and (c) rank-2 tensor representation (tensor).}
    \label{fig:PAPER1_MSE_vs_iterations_allreps}
\end{figure}

Figure \ref{fig:PAPER1_MSE_vs_iterations_allreps} shows these errors as a function of the number of Landweber iterations used to compute the filters. For less than 25 Landweber iterations, the MSE between the algebraic filters and iterative reconstructions is at least 5 times smaller than the MSE between the iterative reconstruction and the ground truth across all object representations. As the number of iterations increases, this gap is gradually bridged due to the accrual of inaccuracies in the filter computations from the shift-equivariance approximation. This result highlights the fidelity of the proposed method's approximation to a given iterative reconstruction. {For} iteration counts commonly used in practice {(around 50 iterations)}, the algebraic filters remain well within the intrinsic reconstruction error of the iterative method itself, indicating that the approximation is sufficiently accurate for practical purposes. 
{The approximation accuracy could potentially be improved by introducing a supersampled ray geometry~\cite{Pelt2015}, improving the shift-equivariance of $\mathbf{\overline{W}}$. While this approach would increase the computational cost of the pre-computation phase, the actual reconstruction time using the resulting filters would remain unaffected.}

\begin{figure}
    \centering
    \includegraphics[width=0.8\linewidth]{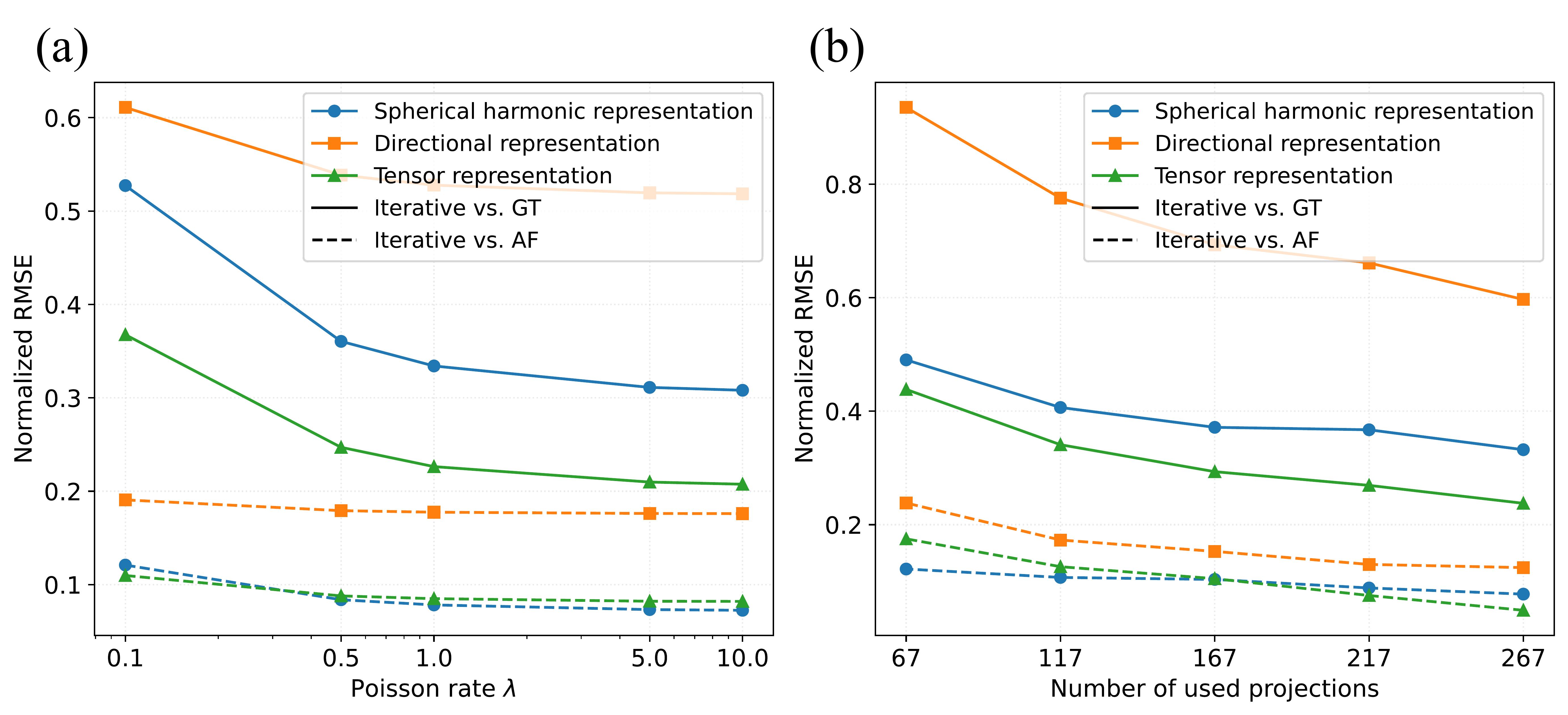}
    \caption{{Robustness of the algebraic filters reconstruction against measurement noise and reduced angular sampling, for $k=50$ Landweber iterations and the three considered scattering tensor representations. (a) Root-mean-square error (RMSE) normalized to the standard deviation of the corresponding representation's synthetic ground truth, as a function of the Poisson noise rate parameter (a lower $\lambda$ simulates higher noise). (b) Normalized RMSE as a function of the number of tomographic projections. Solid lines indicate the error of the Landweber reconstruction ('Iterative') relative to the ground truth ('GT'), and dashed lines represent the error between the algebraic filters reconstruction ('AF') and the Landweber reconstruction.}}
    \label{fig:PAPER1_M_nrmse}
\end{figure}

{To assess the robustness of the proposed method under non-ideal acquisition conditions, the simulation study was extended to include varying levels of measurement noise and reduced angular sampling. Poisson noise was applied the simulated projection data via a rate parameter $\lambda$, and reconstructions were performed with $k=50$ Landweber iterations. As shown in Fig. \ref{fig:PAPER1_M_nrmse}(a), the Landweber reconstruction error relative to the ground truth (solid lines) increases for all representations as the noise level rises. However, the approximation error between the algebraic filters method and the iterative baseline (dashed lines) remains low and stable. Fig. \ref{fig:PAPER1_M_nrmse}(b) illustrates a similar robustness against limited angular sampling, where various numbers of equally-spaced tomographic views are removed from the full geometry.}

\begin{figure}
    \centering
    \includegraphics[width=0.8\linewidth]{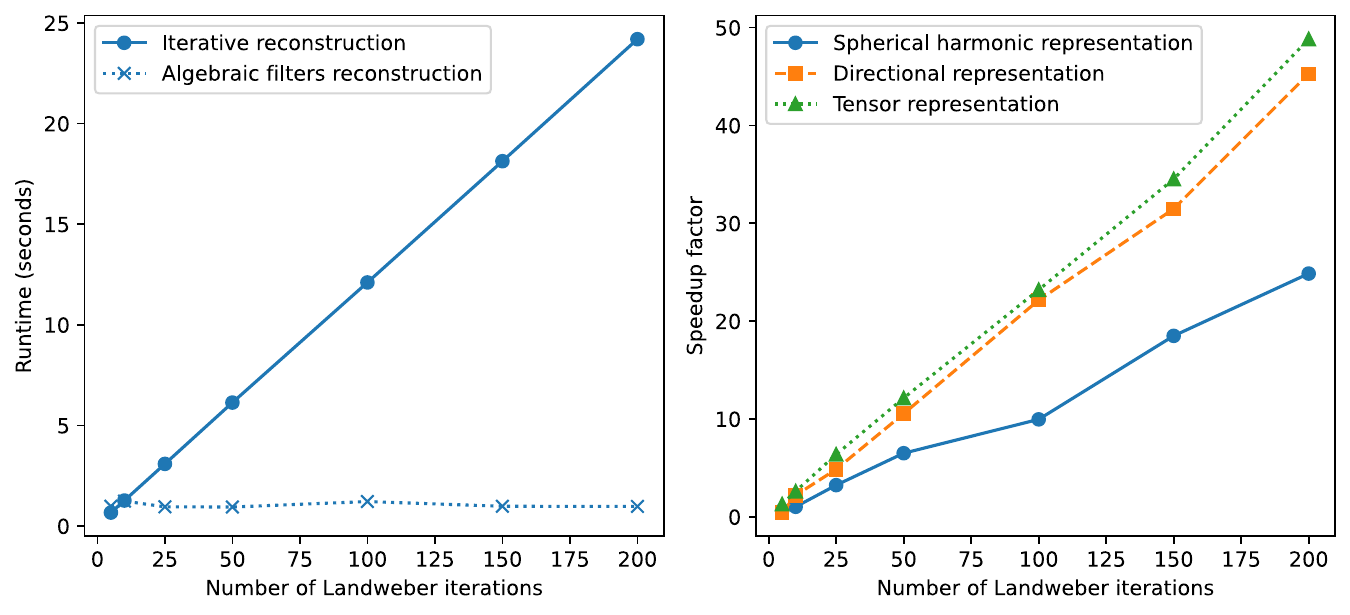}
    \caption{(a) Reconstruction runtimes of the Landweber iterative and algebraic filters methods for the spherical harmonic representation of the scattering tensor, shown as a function of the number of iterations. (b) Speed-up factors of the algebraic filter method relative to iterative reconstruction for all three object representations considered.}
    \label{fig:PAPER1_speedup_dualplot}
\end{figure}

Fig. \ref{fig:PAPER1_speedup_dualplot} shows the runtimes of both reconstruction methods for the spherical harmonic representation, as well as the speed-up factors for all three representations depending on the number of Landweber iterations, which is defined as the ratio between the runtimes of the iterative and the algebraic filter reconstructions.
While the runtime of the iterative method increases linearly with the number of iterations, the algebraic filter reconstruction requires only a single filtering and back-projection operation, resulting in a nearly constant runtime. In most cases, the proposed method achieves around an order of magnitude of acceleration, with speed-ups increasing with the number of iterations used as reference for the Landweber reconstruction. {Additionally, similarly to conventional FBP, the presented reconstruction method is easily parallelizable due to its per-voxel formulation. This aspect implies that the speed-ups of Fig. \ref{fig:PAPER1_speedup_dualplot} scale linearly with the volume size for a given representation.}

{As observed in Fig. \ref{fig:PAPER1_speedup_dualplot}(b), the relative speed-up varies across the different object representations. This variation arises from the complex interplay between the number of reconstructed tensor components $C$, the number of measurement channels $S$, and the model embedded in each specific mixing operator $\mathbf{Y}$. Furthermore, the parallelization used in our implementation may favour certain representations due to better alignment between the available compute threads and the number of required filtering operations. The relative runtime advantage might therefore decrease with an increase in $C$ (or $S$), such as in texture tomography, where the object may be represented with a large number of basis functions \cite{Carlsen2024}. A comprehensive characterization and optimization of these representation-specific nuances remains an interesting avenue for future research.}

The results of this simulation study show that the algebraic filters provide an accurate approximation of iterative reconstructions at a fraction of the computational cost, and that this feature holds across the considered different tensor representations and corresponding scattering models. These advantages persist for the geometry the filters were computed for: they can be reused without modification for any sample, highlighting the method's potential for applications requiring high-throughput or real-time reconstruction.





\subsection{Application to experimental datasets}


To evaluate the method under real measurement conditions, we apply the same reconstruction procedure to three experimental datasets: one for each of the modalities depicted in figure \ref{fig:PAPER1_diagrams_modalities}. The first dataset corresponds to a SASTT scan (Fig. \ref{fig:PAPER1_diagrams_modalities}(a)) of a trabecular bone from a human vertebra \cite{Liebi2015, trabecular_dataset}; the second originates from a GITT scan (Fig. \ref{fig:PAPER1_diagrams_modalities}(b)) of orthogonally oriented carbon fibre bundles embedded in a PMMA box \cite{Kim2020,pmma_dataset}; and the third dataset originates from a SBTT scan (Fig. \ref{fig:PAPER1_diagrams_modalities}(c)) of a sample composed of four carbon fibre rods glued together in different orientations \cite{Lautizi2024b}. In order to preserve the integrity of the physical modelling that precedes each experimental modality, we use the spherical harmonic representation for reconstructing the first dataset, the directional sampling representation for the second, and the symmetric rank-2 tensor representation for the third dataset.

\begin{figure}
    \centering
    \includegraphics[width=\linewidth]{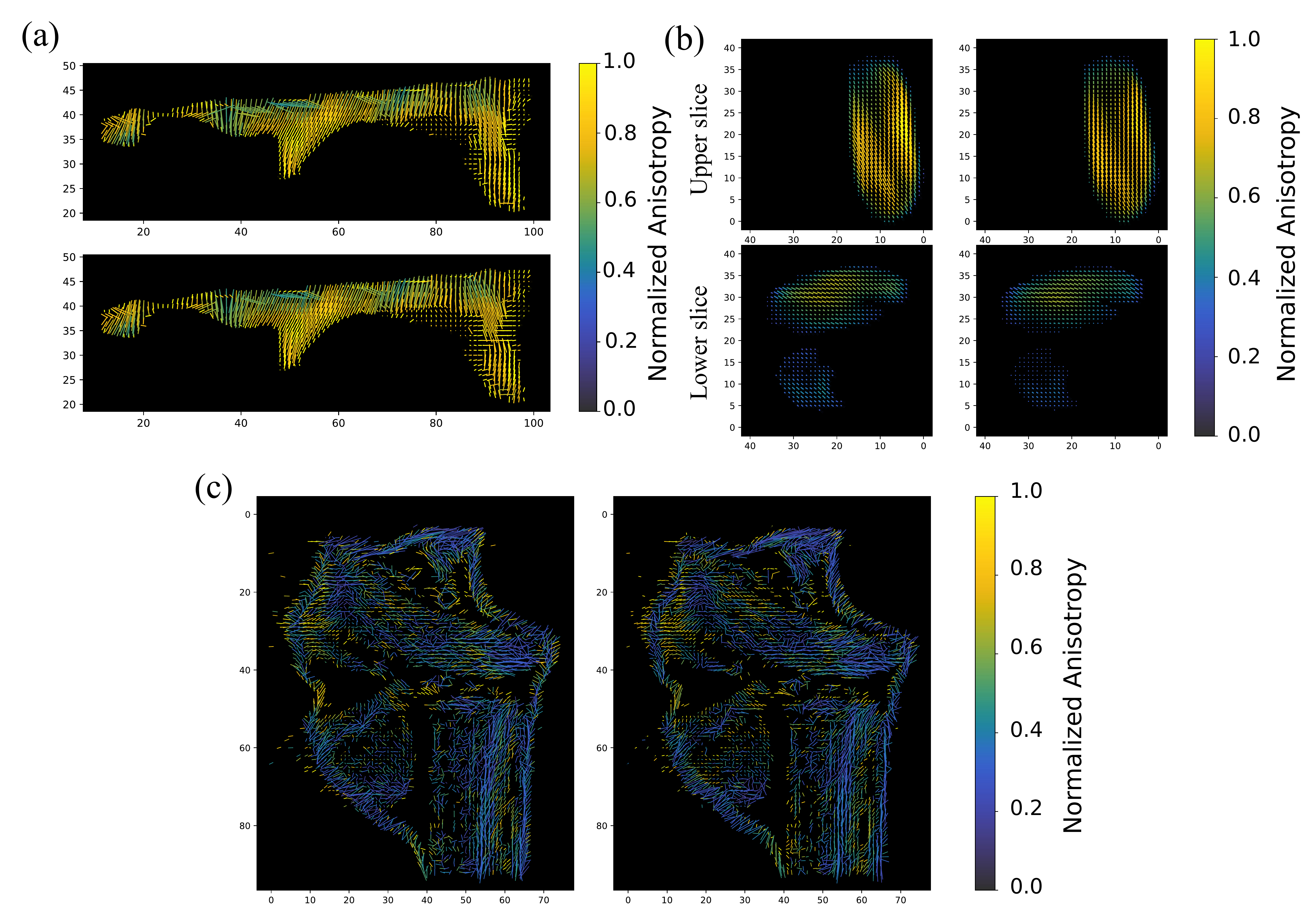}
    \caption{Reduced tensor visualization of slices through reconstructions of real-world X-ray scattering-based tensor tomography datasets to qualitatively assess the fidelity of the algebraic filters reconstruction to its iterative counterpart. Three reduced quantities are represented: the average scattering strength is visualized as the size of the plotted sticks; the main scattering direction in the plane of the slice is represented directly as their orientation; and the normalized anisotropy of the scattering (relating to the degree of alignment of the scatterers) is visualised as their colour. (a) Slice through the reconstruction of a human trabecula \cite{Liebi2015, trabecular_dataset} acquired within the SASTT modality (Fig. \ref{fig:PAPER1_diagrams_modalities}(a)) (b) Two different-height slices of the reconstruction of a PMMA box with carbon fibre bundles \cite{Kim2020,pmma_dataset} resulting from an acquisition within the GITT modality (Fig. \ref{fig:PAPER1_diagrams_modalities}(b)). (c) Slice through the reconstruction of a sample composed of four carbon fibre rods glued in different orientations \cite{Lautizi2024b}, acquired using the SBTT modality (Fig. \ref{fig:PAPER1_diagrams_modalities}(c)).}
    \label{fig:PAPER1_realdata_reconstructions_2}
\end{figure}

The reconstructions are carried out using both the Landweber method and the algebraic filters approach for $k=50$ iterations, using the appropriate mixing operator for each representation. Figure \ref{fig:PAPER1_realdata_reconstructions_2} shows reduced tensor visualizations of representative slices from all reconstructions. The algebraic filter reconstructions preserve the main scattering strength, dominant orientation, and degree of anisotropy observed in the iterative reconstructions. Differences between the two are minimal and localized, consistent with the behaviour observed in simulation. Most importantly, no degradation or instability is observed when transitioning from simulated to real data, despite differences in the physics of imaging and measurement noise.

{To quantify the quality of the reconstructions for the three experimental datasets, we computed the per-voxel relative scattering strength error and the angular misalignment between the dominant scattering angle of the algebraic filters method and the Landweber baseline. Results are summarized in Table \ref{tab:PAPER1_realdata_metrics}. Because the dominant angle is ill-defined in regions where the scattering strength is small, angle misalignments were weighted by the scattering strength of the Landweber reconstruction. For the GITT and SASTT datasets, the algebraic filters method yielded a median angular error of 4.86° and 2.23° and a relative strength error of 3.84\% and 7.46\%, respectively. For the SBTT dataset, higher median angular error (20.84°) and strength error (14.09\%) were observed, potentially caused by differences in sample complexity and noise characteristics between the experimental acquisitions.}

\begin{table}[htbp]
\caption{{Statistics of the distribution of misalignments between the dominant angles of the algebraic filters and Landweber reconstructions of Fig. \ref{fig:PAPER1_realdata_reconstructions_2}, weighted by the corresponding scattering strength of the baseline reconstruction. The average relative scattering strength MSE between the two reconstructions is also presented. These figures are presented for the three experimental datasets relative to the SASTT, GITT and SBTT modalities considered.}}
  \label{tab:PAPER1_realdata_metrics}
  \centering
\begin{tabular}{ccccc}
\hline
 & \multicolumn{3}{c}{Angle misalignment} & \\
\cline{2-4}
Dataset & Mean ($^\circ$) & Median ($^\circ$) & Std. dev. ($^\circ$) & Rel.\ strength MSE (\%) \\
\hline
SASTT &  8.10 &  2.23 & 20.30 &  7.46 \\
GITT  &  5.86 &  4.86 &  4.17 &  3.84 \\
SBTT  & 27.42 & 20.84 & 21.06 & 14.09 \\
\hline
\end{tabular}
\end{table}

Overall, the experimental results confirm the conclusions drawn from the simulation study: the algebraic filters provide a stable and accurate approximation of iterative reconstructions at a lower computational cost, and do so consistently across experimental modalities. {The robustness of our approach to experimental miscalibrations such as angle misalignments or detector shifts remains subject to further research, although earlier research has indicated that filter-based methods could be more robust than iterative methods~\cite{Pan2009}.}



\section{Conclusion} \label{sec:PAPER1_Conclusion}

We have presented a fast reconstruction method for X-ray scattering-based tensor tomography based on the computation of algebraic filters that approximate the output of linear iterative reconstruction algorithms. By explicitly separating the conventional tomographic projector from the view-dependent mixing operator, the approach can be applied to different tensor representations and {parallel-beam} experimental modalities without requiring changes to the underlying framework.

Using both simulated data and three experimental datasets from scanning small-angle X-ray scattering tensor tomography and grating- and speckle-based dark-field tensor tomography, we demonstrated that the proposed method \review{can provide} accurate approximations of iterative Landweber reconstructions while reducing computation times by at least an order of magnitude {across different representations} of the scattering tensor and real-world measurement conditions. Once computed for a given acquisition geometry, the filters can be reused without modification, enabling high-throughput and real-time reconstruction workflows.

The presented results show that this direct, fast reconstruction method can replace {linear, regularization-free} iterative reconstructions {in a representation-independent manner that} \review{preserves overall reconstruction quality, although variations in angular and strength fidelity can occur depending on the specific tensor representation, experimental modality, and sample complexity (as observed with SBTT)}. This advancement in the context of X-ray tensor tomography opens up possibilities for in-line 3D scattering-based imaging and real-time reconstruction for dynamic acquisition protocol design {across different acquisition modalities}. Future work may explore filter design for non-linear iterative methods, as well as local tensor tomography.

\begin{backmatter}
\bmsection{Funding}
The authors acknowledge funding from Horizon Europe through the MSCA Doctoral Network RELIANCE, grant no. 101073040.

\bmsection{Acknowledgment}
The authors would like to thank C. Gutiérrez Bolaños, S. Wang and M. Liebi for insightful discussions about the SASTT modality and reconstruction using the \textit{mumott} library; J. Kim, for the explanations about the original implementation of the algebraic filters method for GITT and for conceding a related experimental dataset; and G. Lautizi, V. Di Trapani and P. Thibault for kindly providing an experimental speckle-based dark-field tensor tomography dataset.

\bmsection{Disclosures}
The authors declare no conflicts of interest.










\bmsection{Data availability} The artificial 'M'-shaped tensor phantom is available in Ref. \cite{M_dataset}. The experimental trabecular bone SASTT dataset and the PMMA box with carbon fibre bundles dataset are both publicly available in Ref. \cite{trabecular_dataset} and Ref. \cite{pmma_dataset}, respectively. The remaining experimental carbon fibre rods dataset used in Ref. \cite{Lautizi2024b}, of which we do not have ownership, is not currently publicly available. A set of notebooks containing a complete implementation of representation changes, forward models, and of the proposed algorithm for the simulated tensor phantom will be made available upon publication.






\end{backmatter}




\bibliography{af_paper}

\end{document}